\documentclass[twocolumn,showpacs,preprintnumbers,amsmath,amssymb]{revtex4}
\usepackage{graphicx}
\usepackage{dcolumn}
\usepackage{bm}

\begin{document}

%\preprint{APS/123-QED}
\title{Towards surface quantum optics with Bose-Einstein condensates in evanescent waves}

\author{Helmar Bender}
\author{Philippe Courteille}
\author{Claus Zimmermann}
\author{Sebastian Slama}
\affiliation{Physikalisches Institut, Eberhard-Karls-Universit\"at T\"ubingen,\\
Auf der Morgenstelle 14, D-72076 T\"ubingen, Germany}

\date{\today}

\begin{abstract}
We present a surface trap which allows for studying the coherent
interaction of ultracold atoms with evanescent waves. The trap combines a
magnetic Joffe trap with a repulsive evanescent dipole potential. The
position of the magnetic trap can be controlled with high precision
which makes it possible to move ultracold atoms to the
surface of a glass prism in a controlled way. The optical potential
of the evanescent wave compensates for the strong
attractive van der Waals forces and
generates a potential barrier at only a few hundred nanometers from
the surface. The trap is tested with $^{87}$Rb Bose-Einstein
condensates (BEC), which are stably positioned at distances from the surfaces below
one micrometer.
\end{abstract}


\maketitle

The interaction of ultracold atoms with surfaces has attracted much
attention in the past for various technological and fundamental
reasons. By the use of repulsive evanescent waves (EW)
at a prism surface atom mirrors were constructed, from which both
atomic beams and cold atomic clouds were reflected \cite{Balykin87,Kasevich90}.
EWs were also used for generating steep surface traps in which two-dimensional
Bose-Einstein condensates can be prepared \cite{Rychtarik04}.
Atom-surface interactions are also interesting for investigating fundamental aspects
of quantum electrodynamics. Close to surfaces the
electromagnetic vacuum fluctuations lead to the emergence of surface
potentials like van der Waals or Casimir-Polder potentials. These
potentials have been measured with hot atomic beams
\cite{Sandoghdar92}, in vapor nanocells \cite{Fichet07}, in atomic
mirrors \cite{Landragin96}, by quantum reflection \cite{Shimizu01,
Druzhinina03,Pasquini04} and in magnetic traps \cite{Obrecht07}.\\

    \begin{figure}[ht]
        \centerline{\scalebox{0.7}{\includegraphics{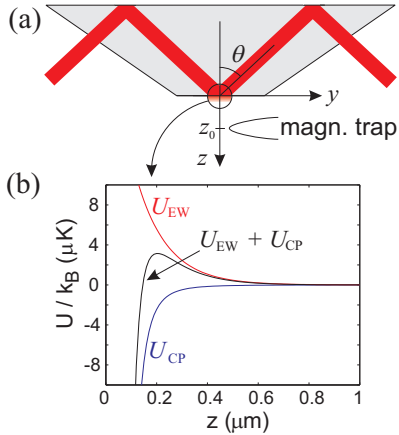}}}\caption{
        (a) Schematic view of the prism. The magnetic trap is situated below the prism surface
        at a distance $z_0$ and can be shifted towards the surface, where the Casimir-Polder
        force and the repulsive dipole potential become dominant. This is illustrated in
        (b), where the Casimir-Polder potential $U_{\textrm{CP}}$ and the evanescent wave
        potential $U_{\textrm{EW}}$  add up to form a potential barrier at few hundred nanometers
        from the prism surface.}
        \label{fig:prismscheme}
    \end{figure}

Inspired by the first atom mirror experiments, proposals have been developed for non-destructive detection of atoms
by evanescent waves with the perspective of QND measurements of atomic number fluctuations
\cite{Aspect95,Courtois95}. The idea was that atoms interacting with an EW act as a refractive index
medium and thereby shift the phase of the totally reflected light field. A measurement of this phase
shift should therefore reveal the atom number. By choosing a large detuning of the light frequency from
atomic resonances, the atomic cloud remains undisturbed by this measurement. This proposal is complementary to
experiments with resonant light fields which allow to measure the atom number destructively \cite{Cornelussen02}.
In the situation considered in \cite{Aspect95} atoms are released from
a magneto-optic trap (MOT) and bounce off an atom mirror. A quantitative analysis revealed a very small phase shift
which will be barely detectable. The main problems are the facts that (i) the density of atoms from the MOT is too low and (ii)
the interaction time during the reflection is very short which results in a high shot noise.
Other problems with reflection experiments are random scattering due to surface roughness
\cite{Henkel97, Savalli02, Perrin06}. We are able to circumvent these problems by using
BECs which have four orders of magnitude larger densities and by keeping the atoms in a trap,
which allows for adjusting the interaction time. Calculation with our parameters following the
methods in \cite{Aspect95} result in detectable phase shifts on the order of $\Delta\phi\sim10^{-4}\,\textrm{rad}$ and
signal to shot-noise ratios on the order of a few thousand.\\

The surface trap presented here uniquely combines the advantages of evanescent waves and magnetic fields.
Magnetic traps have the advantage that they can be easily shifted by offset fields such that the atoms can
be brought to the surface in a controlled way. However, magnetic traps are less steep than
dipole traps. Therefore the problem occurs that the magnetic trapping potential is opened by the attractive
Casimir-Polder potential close to surfaces which leads to the loss of atoms \cite{Lin04}. This can be avoided by
an additional dipole potential that is generated by a blue detuned evanescent wave
(Fig.\ref{fig:prismscheme} a). The repulsive dipole potential is able to partially compensate for the
Casimir-Polder potential
thereby producing a potential barrier at only a few hundred nanometers from the surface (Fig.\ref{fig:prismscheme} b).
This potential barrier prevents the atoms from being lost when the magnetic trap is shifted to the
surface.\\

In the following, simulations of our combined surface
trap are shown. The trapping potential is determined by the Casimir-Polder
potential $U_{\textrm{CP}}$, the evanescent wave potential
$U_{\textrm{EW}}$, the magnetic trapping potential
$U_{\textrm{magn}}$ and the gravitational potential $U_{\textrm{g}}$
\begin{equation}
U_{\textrm{tot}}=U_{\textrm{CP}}+U_{\textrm{EW}}+U_{\textrm{magn}}+U_{\textrm{g}}~.
\end{equation}
For the simulation of the surface potential, we use the formula for
the retarded regime
\begin{equation}\label{eq:UCP}
U_{\textrm{CP}}(z)=-\frac{C_4}{z^4},
\end{equation}
with $z$ the distance from the surface and
\begin{equation}
C_4=\frac{1}{4\pi\epsilon_0}\cdot\frac{3\hbar c\alpha(0)}{8\pi}\cdot
\frac{\epsilon(0)-1}{\epsilon(0)+1}\cdot\Phi(\epsilon(0))\,
\end{equation}
where $\epsilon_0$ is the dielectric permittivity of vacuum, $c$ the
speed of light, $\alpha(0)=5.26\cdot10^{-39}\,\textrm{Fm}^2$ the
static polarizability of Rb, $\epsilon(0)=2.25$ the static
dielectric constant of the glass substrate corresponding to a
refractive index of of $n=1.5$ and $\Phi(2.25)=0.29$ is
given by a formula in \cite{Lifshitz61}. From these values we get a
$C_4$ coefficient of $C_4=1.78\cdot10^{-55}\textrm{Jm}^4$. At the
distances relevant for our simulations the formula of the retarded
regime (\ref{eq:UCP}) is a sufficient approximation, although more
precise interpolation formulas exist \cite{Druzhinina03}. The evanescent wave potential
exponentially decays with distance from the surface $z$ and has a
gaussian shape in the transverse direction $x,y$ with beamwaists
$w_{x,y}$:
\begin{equation}\label{eq:UEW}
U_{\textrm{EW}}(x,y,z)=U_0\cdot\exp\left\{-\frac{z}{\lambda_p}-2\frac{x^2}{w_x^2}-2\frac{y^2}{w_y^2}\right\},
\end{equation}
where the penetration depth is given by
$\lambda_p=\left(k\cdot\sqrt{n^2\sin(\theta)^2-1}\right)^{-1}$, with
$k=2\pi/\lambda$ the wavevector of the light with wavelength
$\lambda=765\,\textrm{nm}$ and $\theta$ the incidence angle (see
figure \ref{fig:prismscheme}). The maximum value of the potential at
the surface is given by the dipole potential
\begin{equation}\label{eq:dipolepotential}
U_0=\frac{\pi c^2
\Gamma}{2\omega^3}\cdot\frac{I_{\textrm{ev}}}{\Delta}\,
\end{equation}
with $\Gamma=2\pi\times 6\,\textrm{MHz}$ the natural linewidth of the
Rb 5P state, $\omega$ the frequency of light, $I_{\textrm{ev}}$ the maximum
light intensity in the evanescent wave and $\Delta$ the mean
detuning of the laser frequency from the Rb $D_1$ and $D_2$ line.
The magnetic trap is harmonic
\begin{equation}\label{eq:magnetictrap}
U_{\textrm{magn}}(x,y,z)=\frac{1}{2}m\left(\omega_x^2x^2+\omega_y^2y^2+\omega_z^2(z-z_0)^2\right)\,
\end{equation}
with trap frequencies of $\omega_x\sim2\pi\times\,25\,\textrm{Hz}$,
$\omega_y=\omega_z\sim2\pi\times\,200\,\textrm{Hz}$ and $m$ the atomic
mass. The magnetic field minimum is located at a distance from the
surface $z_0$ which is adjustable in our experiment. In
good approximation the transverse coordinates of the magnetic
field minimum coincide with the position of maximum evanescent light
intensity. Finally the gravitational potential is given by
\begin{equation}\label{eq:magnetictrap}
U_{\textrm{g}}(z)=-mgz\,
\end{equation}
where the negative sign accounts for the fact that the
prism is mounted upside down in the chamber.\\

    \begin{figure}[ht]
        \centerline{\scalebox{0.7}{\includegraphics{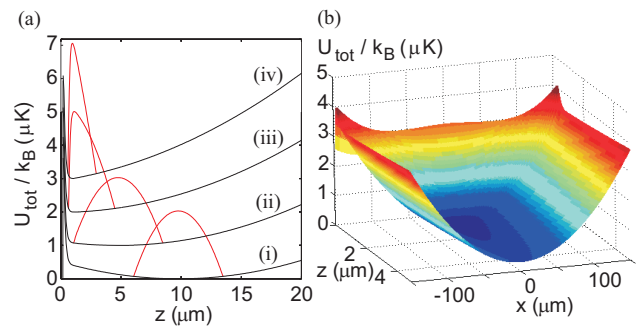}}}\caption{
        (a) Simulated potential and atom density for
        different positions of the magnetic minimum $z_0$: (i) $z_0=0\,\mu\textrm{m}$,
        (ii) $z_0=-5\,\mu\textrm{m}$, (iii) $z_0=-10\,\mu\textrm{m}$ and (iv) $z_0=-15\,\mu\textrm{m}$.
        The curves are shifted relative to each other by $1\,\mu\textrm{K}$ for
        better visibility.
        In these simulations the laser power is $P=500\,\textrm{mW}$, the reflection angle is $\theta=47.5^\circ$
        corresponding to a penetration depth of $\lambda_p=243\,\textrm{nm}$ and the beamwaists
        are $w_x=170\,\mu\textrm{m}$ and $w_y=240\,\mu\textrm{m}$.
        (b) Two-dimensional plot of the simulation result corresponding
        to curve (iv) in fig. (a). The atomic density is omitted here for clarity.}
        \label{fig:trapsimulation}
    \end{figure}

Figure \ref{fig:trapsimulation} (a) shows simulations
of realistic trapping potentials for different positions of the
magnetic field minimum $z_0$. Negative values of $z_0$ correspond to
magnetic field minima behind the prism surface. The actual position of the
trap though is at positive $z$ values due to the combination with
the other potentials. In addition, the atomic density in Thomas Fermi approximation is plotted for $N=10^5$
atoms. By shifting the trap towards the surface the atomic cloud is
compressed. Fig. \ref{fig:trapsimulation} (a) shows a cut through the center
of the trap along the $z$-axis. Fig. \ref{fig:trapsimulation} (b)
shows the simulated potential as a function of both $z$ and $x$
axis. A small reduction of the trap depth due to two saddle points is
observed at about $x=\pm70\,\mu\textrm{m}$, because of the
transversal gaussian shape of the repulsive dipole potential.\\

    \begin{figure}[ht]
        \centerline{\scalebox{0.6}{\includegraphics{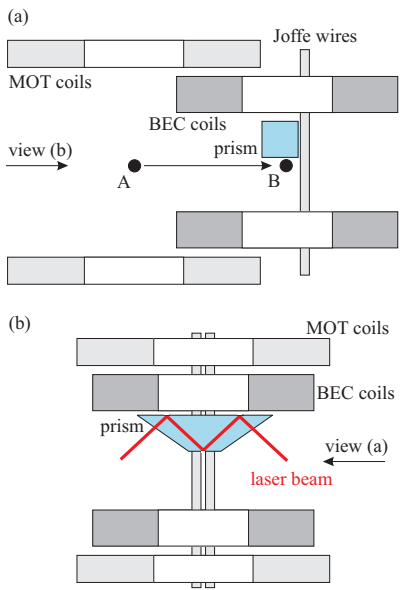}}}\caption{
        (a) and (b) Schematic side views of the experimental setup. The magneto-optic
        trap is produced at position A. After adiabatic transfer to position B the atoms
        are evaporatively cooled in a Joffe-Pritchard type of wire trap, which
        is located about $100\,\mu\textrm{m}$ below the
        surface of a glass prism. The Bose-condensed atoms are shifted
        in the magnetic trap to the surface. After the interaction they are
        withdrawn from the surface and then let adiabatically expanded for
        absorption imaging. }
        \label{fig:setup}
    \end{figure}

The schematic setup of our experiment is sketched in Figure
\ref{fig:setup}. At position A $^{87}$Rb atoms are collected in a
magneto-optic trap. After loading into a purely magnetic trap the atoms are
adiabatically transferred to a Joffe-Pritchard type of wire trap which is formed by a
second pair of coils (BEC coils) and two wires which are running
vertically through the coils. There the atoms are further cooled by
forced evaporation. Within this trap we can produce Bose-Einstein
condensates with up to approximately $6\cdot 10^5$ atoms. The
special feature of this wire trap is the fact that it can be easily shifted
vertically by applying an
external offset field. By this we are able to move the atoms very
close to the surface of a glass prism which is mounted at the upper
magnetic coil. The prism being attached upside down enables us to
analyze the atomic cloud after the interaction with the prism
surface in time of flight (TOF) absorption imaging.\\
  \begin{figure}[ht]
        \centerline{\scalebox{0.9}{\includegraphics{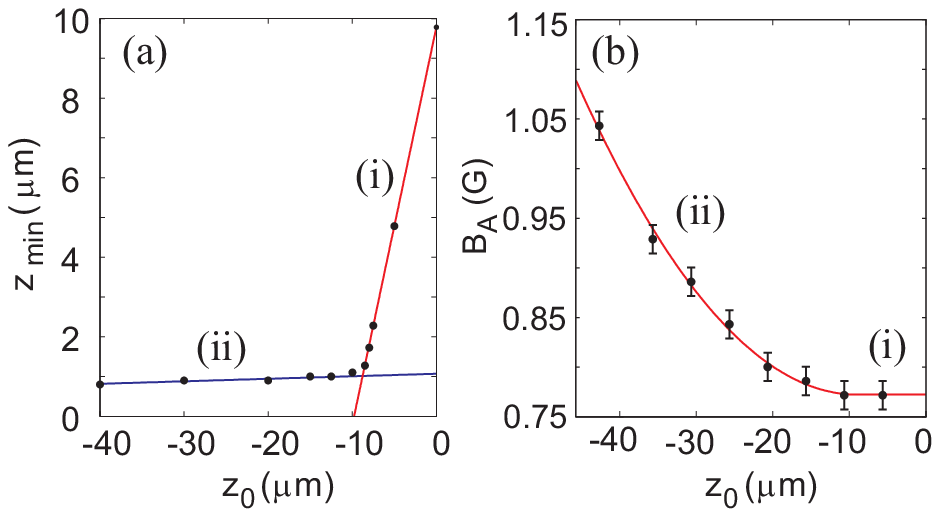}}}\caption{
        (a) Simulated position of the trapping minimum $z_{\textrm{min}}$ as a function
        of the position of the magnetic field minimum $z_0$. Negative values of $z_0$ correspond
        to a magnetic field minimum behind the prism surface. The simulated data points are fitted by two
        linear curves corresponding to the two regimes, where (i) the trap is moving along with the
        magnetic trap and (ii) the trap is held back by the potential barrier at the surface. (b) Measured magnetic field
        $B_{\textrm{A}}$ at the position of the atoms as a function of $z_0$. In regime (i) $B_{\textrm{A}}$
        stays constant, whereas in regime (ii) $B_{\textrm{A}}$ is rising quadratically. Errorbars
        are due to the uncertainty in the measured radio frequency.}
        \label{fig:zshift}
    \end{figure}

The position of the trap
$z_{\textrm{min}}$ as a function of the position of the magnetic
field minimum $z_{\textrm{0}}$ is illustrated in
Fig. \ref{fig:zshift} (a). At distances of more than about one
micrometer from the surface, the trap shifts with the magnetic field minimum. This is consistent with
a linear fit of the simulated data points in regime (i) in figure
\ref{fig:zshift} (a) which shows a gradient of
$\frac{dz_{\textrm{min}}}{dz_{\textrm{0}}}=1$. The offset
of $10\,\mu\textrm{m}$ is due to gravitational sagging.
At distances to the surface of below about $1\,\mu\textrm{m}$ the atoms
cannot follow the magnetic potential any more, but are
held back by the evanescent wave potential barrier (regime (ii)).
Only a small shift remains on the order of below
$\frac{dz_{\textrm{min}}}{dz_{\textrm{0}}}\sim 0.01$. The magnetic
field at the position of the atoms $B_\textrm{A}$ therefore increases. We have
measured this increase with radio-frequency (rf) spectroscopy. For each
position of the magnetic field minimum the radio frequency is detuned at which the atoms
are resonantly coupled to an untrapped Zeeman state and thus lost from the trap. The transition
frequency is given by the magnetic field $B$ via
$\hbar\omega_{\textrm{rf}}=g_\textrm{F}\Delta m_{\textrm{F}}\mu_\textrm{B}B$, with
$g_\textrm{F}$ the Land\'e factor, $\Delta m_\textrm{F}$ the difference in the
magnetic quantum number and $\mu_\textrm{B}$ the Bohr magneton. By determining the minimum
radio frequency at which atom losses occur we get the magnetic field
at the bottom of the trap. The result is shown in Fig. \ref{fig:zshift} (b). As we ramp the currents
in the magnetic coils and thereby shift the magnetic trap towards the prism, we observe the transition from
regime (i) where the magnetic offset field is constant to regime (ii) where it grows quadratically.
A fit to the data reveals a magnetic trapping frequency of
$\omega_z=2\pi\times195\,\textrm{Hz}$, which is in good agreement with the value observed in
trap oscillations. We have calibrated the relative shift of the magnetic field minimum to the currents
in the magnetic coils by absorption imaging at larger distances from the surface. The absolute skale is fixed
by the position where the magnetic offset field starts to rise quadratically. This position is taken from the simulation
in Figure \ref{fig:zshift} (a). As it turns out to be very robust against changes in the simulation
parameters, we assume that the absolute position of the magnetic field minimum is calibrated to an error of
approximately $1\,\mu\textrm{m}$.\\

  \begin{figure}[ht]
        \centerline{\scalebox{0.8}{\includegraphics{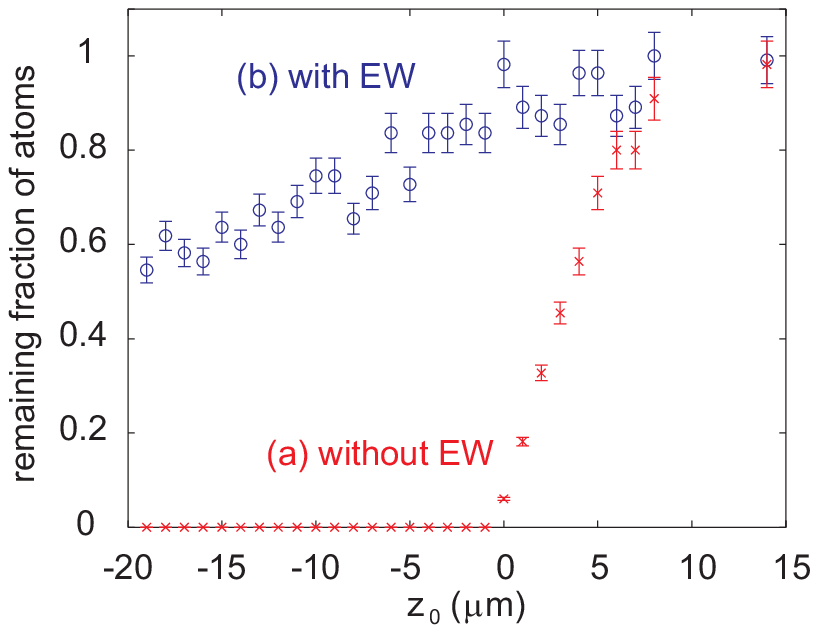}}}\caption{
        Remaining fraction of atoms in the surface trap as a function of the magnetic field minimum $z_0$. We observe
        in curve (a) that without the potential barrier of the evanescent wave, all atoms are lost
        at a distance of approximately $z_{\textrm{min}}=10\,\mu\textrm{m}$. With the evanescent wave
        (curve (b)) atom losses are strongly reduced such that the atoms can be positioned at distances from the surface
        below one micrometer. The errorbars are due to an uncertainty in the measured atom number of 10\%.}
        \label{fig:atomloss}
    \end{figure}

With the help of the evanescent wave the atom losses occurring
in magnetic traps close to surfaces are strongly reduced (Fig. 5) \cite{Lin04}. Here,
a Bose-Einstein condensate is initially prepared in a magnetic trap at
about $z_{\textrm{min}}=40\,\mu\textrm{m}$ below the prism surface. Then the magnetic trap is shifted in
$\tau=200\,\textrm{ms}$
towards the prism surface to a distance $z_0$ of the magnetic field minimum. In order to shift the atoms smoothly
the corresponding coil currents are ramped with a sinusoidal time dependance $I=I_0+\Delta I\cdot\sin(\pi t/\tau)^2$.
After the magnetic trap is immediately shifted back to its starting position within $100\,\textrm{ms}$, the atoms are
released and counted. The experimental results are shown in Figure
\ref{fig:atomloss}. Without the evanescent wave (curve a) the atoms are
lost as soon as the Casimir-Polder potential opens up the magnetic trap. With the given magnetic trapping frequencies
this occurs at a distance of the atoms from the surface of approximately $10\,\mu\textrm{m}$. The observed
width of the decrease of atoms can be well explained by an energy spread given by the chemical potential
of the BEC. However, with the evanescent dipole potential (curve b), the atoms can be shifted much
closer to the surface. The position of the magnetic field minimum of several $10\,\mu\textrm{m}$ behind the surface
corresponds to a distance of the atoms from the surface of below $1\,\mu\textrm{m}$. At such small distances about
half of the atoms are lost within the measuring time. This loss rate can be further reduced by increasing
the laser intensity and thereby the potential barrier. We therefore attribute the atom loss to (i) the evaporation
of atoms over the barrier and (ii) tunneling of atoms through the barrier.\\

We have described a surface trap which combines the flexibility of magnetic traps with the high
gradients of evanescent dipole potentials. In such a trap atoms can be stably brought to the surface of
transparent media with distances below $1\,\mu\textrm{m}$, which is the range of evanescent waves and surface
potentials. This is the key for various experiments with cold atoms in which surface properties are
investigated and exploited. Important examples are dispersive potentials due to van der Waals and Casimir forces
\cite{Kallush05}, but also the fascinating idea of investigating bound surface-atom states by laser-induced
quantum adsorption \cite{Passerat06, Afanasiev07}, which is the analogue to photo-association of atoms.
The crucial question herein
is if atoms can be trapped in these highly excited bound states and if these traps can be tailored by a suitable
structuring of the surface. In this context surface plasmon polaritons which can be used to locally enhance evanescent
waves above surfaces are very promising \cite{Righini07}.\\

We acknowledge financial support by the EuroQUASAR program of the European Science Foundation and by  Projektf\"orderung
f\"ur Nachwuchswissenschaftler der Universit\"at T\"ubingen. We would like to thank Robin Kaiser and
Andreas G\"unther for helpful discussions.

\end{document}